\documentclass[showpacs]{revtex4}

\usepackage{amsmath}
\usepackage{amsfonts}
\usepackage{graphicx}

\begin{document}

\draft
\title{Near field in quantum electrodynamics: Green functions,
Lorentz condition, "nonlocality in the small", frustrated total
reflection}

\author{Mark E. Perel'man}
\affiliation{The Racah Institute of Physics, The Hebrew University
of Jerusalem, 91904 Jerusalem, Israel}

\begin{abstract}

Investigation of near field of QED requires the refuse from an averaging of
the Lorentz condition that smooths out some field peculiarities. Instead of
it Schwinger decomposition of the 4-potential with the Bogoliubov method of
interaction switching in time and in space regions is considered. At such
approach near field is describable by the part of covariant Green function
of QED, the fast-damping Schwinger function formed by longitudinal and
scalar components of $A_{\mu }$ none restricted by light cone. This
description reveals possibility of superluminal phenomena within the near
field zone as a "nonlocality in the small". Some specification of Bogoliubov
method allows, as examples, descriptions of near fields of point-like charge
and at FTIR phenomena. Precisely such possibilities of nonlocal interactions
are revealed in the common QED expressions for the Van-der-Waals and Casimir
interactions and in the F\"{o}rster law.

\end{abstract}

\pacs{03.30.+p, 12.20.-m, 13.40.-f, 68.37.Uv}
\date{\today}
\maketitle

\textbf{Key words:} Lorentz condition, near field, propagators,
superluminal, FTIR.

\section{Introduction}

The usage of 4-component vector $A_{\mu }$ for the field
quantization, when for the Maxwell equations in vacuum only two
components are needed, induced serious problems at early stages of
QED. These problems were initially obviated by Fermi via averaging
the Lorentz condition as $\partial A_{\mu }/\partial x_{\mu
}\left\vert 0\right\rangle =0$ \ [1]. This way had \ led to the
indefinite metrics that formally exhausts the problem of exception
of superfluous components of $A_{\mu }$ by an artificial their
averaging and leaves the theory completely local, although the used
procedure has not a direct physical sense, e.g. [2]. It excludes
also near field from consideration, but at that time its effects
were out of common interests.

With the discovery of Aharonov-Bohm effect [3] the complexity of real
situation could already become more evident: it is impossible to simply cast
away two additional components of 4-potential, as the $\mathbf{E}$ and $%
\mathbf{B}$ do not describe all features of electromagnetic fields. But
moreover, the development of optics into recent decades show greater
complexity of near field than was commonly accepted. The most unexpected
peculiarities represent the phenomena of interference in the scope of near
field (e.g. [4, 5] and references therein) and numerous observations of
superluminal signalling (the reviews [6]), which also may be related with
near fields. All it requires a returning to general problems of near field
and to a revision of common representations for description of these new
phenomena, at that possibly of nonlocal nature.

But can exist nonlocal solutions of covariant equations for little
distances? If to search solutions of the wave equation as $f(x)=f(\mathbf{v}%
t-\mathbf{r})$, then the non-fading solution exists only for $v\leq c$. This
peculiarity strongly forbids superluminal movement as an effective
nonlocality on arbitrary distances. However, this \textit{equation does not
forbid} propagation of faster-than-$c$ "perturbations" on short distances,
of order of uncertainty principle distances, at condition of their quick
damping.

In our article [7] has been established, in the frame of general
phenomenological approach, that the transferring of excitation within near
field on the distance $\Delta x$ are possible only and only as the instant
tunneling, when these distances are inverse relatively energy deficiency to
the nearest stable or resonant (quasistable) state $\Delta \hbar \omega $:

\begin{equation}
\Delta x\cdot \Delta \omega \sim \pi c.\   \label{(1)}
\end{equation}

It defines a "nonlocality in the small" of near field (this "nonlocality" is
restrained by the especial form of the energy-time duration uncertainty
principle, such additional proof of (1) is given in [8]). Thus the
superluminal phenomena can occur, for example, in the domain of anomalous
dispersion or at the frustrated total internal reflection (FTIR). And it is
one of peculiarities of near field that is not evident at usual approach.

Notice, that the common description of field is related to far fields only:
as must be especially underlined, the features of locality were established
and many times checked for far fields only. But the locality of near fields
never was experimentally checked and possibility of their general or at
special conditions nonlocality can not be a priori excluded.

With the purpose of investigating such possibilities we shall try to examine
the structure of near fields without averaging of the Lorentz condition.
Instead of it the Schwinger method of $A_{\mu }$decomposition on transverse,
longitudinal and scalar components will be used [9]. These procedures
(Section 3) will lead to revealing the existence of nonlocal terms within
the known decomposition of covariant Green functions (Section 2).

Then the most general procedure of interaction on and off switching, the
Bogoliubov method [10] for description of QED interactions without the
averaged additional condition, will be used (Section 4). It completes the
construction of near field in absence averaging procedure and shows its
possible nonlocality, manifestations of which quickly disappear with
distance.

In the subsequent Sections we try to apply this approach to some real
phenomena. It requires special partial specializations of the Bogoliubov
method suggested in the Appendix, but contains some arbitrariness' and
therefore subsequent considerations have rather a hypothetical character.

In the Section 5 a near field of point-like charge will be considered with
estimation of its characteristic energies at different distances. In the
Section 6 on some similar base the phenomena of FTIR are considered, which
reveals the dependence of wave numbers values on distances from the surface
and corresponding possibility of interference picture formation leading to
effects of near field optics.

Then it will be shown that the Van-der-Waals forces, in the form deduced in
[11], and the F\"{o}rster law of excitation transfer on short distances are
describable in the frame of QED via near-field nonlocal interactions. The
revealing of such description means that near-field effects are not so
exotic as it can seems and that they are responsible for some known
phenomena.

The results of executed examinations, which can be considered as some
detalisation of the general results of [7, 8], are summed in the
Conclusions, where some further perspectives of the offered method are
mentioned.

\section{Decomposition of canonical Green functions}

Let's consider the covariant Pauli-Jordan function $D(t,\mathbf{r})$
in the variant of Coulomb gauge suggested by Dzjaloshinski and
Pitayevski [11]

\begin{center}
\begin{equation}
D_{ij}(\omega ,\mathbf{r})=(\delta _{ij}+\partial _{i}\partial _{j}/\omega
^{2})D(\omega ,\ \mathbf{r});\qquad D_{i0}=D_{00}=0.  \label{2}
\end{equation}%
\ \ \ \ \ \ \ \ \ \ \ \ \ \ \ \ \ \ \ \
\end{center}

After differentiation it leads to the representation of Green functions of
wave equation ([12], cf. [5],) as

\begin{center}
\begin{equation}
D_{ij}(\omega ,\ \mathbf{r})=\{(\delta _{ij}+e_{i}e_{j})-\frac{i}{\omega r}%
P_{ij}\cot (\omega r)+\frac{1}{(\omega r)^{2}}P_{ij}\}D(\omega ,\mathbf{r}),
\label{3}
\end{equation}
\end{center}

here and below $c=\hbar =1$, $P_{ij}=\delta _{ij}-3e_{i}e_{j}$. (Note that
the decomposition of transfer functions is simpler than decomposition of
field strengths [13].)

Three terms of (3), which correspond to far, intermediate (or transient) and
near fields, are expressed in the $(t,\mathbf{r})$ representation through
the Pauli-Jordan function and the generalized singular function $D_{N}(t,\
\mathbf{r})$($\partial _{x}\equiv \partial /\partial x$):

\begin{center}
\begin{equation}
D_{ij}(t,\ \mathbf{r})|_{FF}=(\delta _{ij}+e_{i}e_{j})D(t,\mathbf{r});
\label{4}
\end{equation}%
\quad

\qquad

\begin{equation}
D_{ij}(t,\ \mathbf{r})|_{IF}=\frac{1}{4\pi r}P_{ij}\theta
(r^{2}-t^{2})\equiv \frac{1}{r}P_{ij}\ \partial _{t}D_{N}(t,\mathbf{r});
\label{5}
\end{equation}%
\quad\

\qquad

\begin{equation}
D_{ij}(t,\ \mathbf{r})|_{NF}=\frac{1}{4\pi r^{2}}P_{ij}\{sgn(t)\theta
(t^{2}-r^{2})+\frac{t}{r}\theta (r^{2}-t^{2})\}\equiv \frac{1}{r^{2}}%
P_{ij}D_{N}(t,\ \mathbf{r}).  \label{6}
\end{equation}%
\quad\
\end{center}

The function $D_{N}(t,r)$, valuable for all subsequent analysis, can be
determined immediately by (6) and in more details will be considered below.

This decomposition shows nonlocality of near field, increasing with time,
and demonstrates that transitions (5) between near and far fields are
concentrated in the space-like domain. It means that near field of any
charge can be nonlocal, i.e. it supposes, in particular, the possibility of
propagation of quickly relaxing superluminal perturbations within near
fields.

Virtual "quanta of near field" and "dressed", i.e. free photons are clearly
distinguishable in the $(\omega ,\mathbf{k})$ representation (here and below
$k=|\mathbf{k}|$):$\qquad $%
\begin{equation}
D_{ij}(\omega ,\mathbf{k})|_{FF}=(\delta _{ij}+e_{i}e_{j})\frac{2}{(2\pi
)^{3}i}\varepsilon (\omega )\delta (\omega ^{2}-k^{2});  \label{4'}
\end{equation}

\begin{center}
\quad

\begin{equation}
D_{ij}(\omega ,\ \mathbf{k})|_{IF}=\frac{1}{4\pi 2i\omega k}P_{ij}\theta
(k^{2}-\omega ^{2});  \label{5'}
\end{equation}%
\qquad

\quad

\begin{equation}
D_{ij}(\omega ,\ \mathbf{k})|_{NF}=\frac{1}{8\pi ^{2}i\omega ^{2}k}%
P_{ij}\{|\omega |\theta (\omega ^{2}-k^{2})+k\theta (k^{2}-\omega ^{2})\}.
\label{6'}
\end{equation}%
\qquad
\end{center}

These expressions clearly show that when quanta of far field are the free
photons with light speed $c$ in vacuum, the speed of virtual quanta
propagation is not restricted. The transitions between them, described by an
intermediate field, are evidently tunnel processes with an energy deficiency
relative to momentum.

Note, that for functions $D(\omega ,\ \mathbf{k})$, $D^{(1)}(\omega ,\
\mathbf{k})$ and their combinations the gauge (2) actually coincides with
the more widespread Coulomb gauge:

\begin{center}
\begin{equation}
D_{ij}(\omega ,\ \mathbf{k})=(\delta _{ij}-k_{i}k_{j}/k^{2})D(\omega ,\
\mathbf{k});\quad D_{00}=-1/k^{2};\quad D_{i0}=0.  \label{7}
\end{equation}%
\quad
\end{center}

The decomposition similar (3) can be executed, certainly, with singular
functions of non-uniform wave equation $D_{c}(\omega ,\mathbf{k})$, etc.

\section{Schwinger decomposition of 4-potential}


For uncovering the origin and physical sense of the function $D_{N}$ some
properties of 4-potential \ construction should be considered without
averaging of its 4-divergence over vacuum, i.e. without usage of the
Fermi-Lorentz condition.

For this aim the 4-vector $A_{\mu }$ or its frequencies parts in any system
of readout must be covariantly \ decomposed by the Schwinger method [9] onto
the far field $A_{\mu }^{(f)}$ and two auxiliary fields, longitudinal $%
\Lambda _{||}$ and scalar (temporal) $\Lambda _{0}$:

\begin{center}
\begin{equation}
A_{\mu }(x)=A_{\mu }^{(f)}(x)+n_{\mu }(n\partial )\Lambda _{0}(x)-(\partial
_{\mu }+n_{\mu }(n\partial ))\Lambda _{||}(x),  \label{8}
\end{equation}%
\

\bigskip

$n_{\mu }$ is the unit time-like vector, $n_{\mu }^{2}=1$,
$(n\partial )\equiv n_{\mu }\partial _{\mu }$. All three fields
independently obey the wave equations:

\begin{equation}
\Box A_{\mu }^{(f)}(x)=\Box \Lambda _{0}(x)=\Box \Lambda _{||}(x)=0.
\label{9}
\end{equation}

\end{center}

\bigskip

Far field $A_{\mu }^{(f)}$ is transverse and satisfies the classical Lorentz
condition:

\begin{center}
\begin{equation}
n_{\mu }A_{\mu }^{(f)}(x)=\partial _{\mu }A_{\mu }^{(f)}(x)=0.  \label{9'}
\end{equation}
\end{center}

Auxiliary fields are defined via $A_{\mu }$:\

\begin{center}
\begin{equation}
\partial _{\mu }A_{\mu }=(n\partial )^{2}(\Lambda _{0}-\Lambda _{||});\qquad
n_{\mu }A_{\mu }=-(n\partial )\Lambda _{0}(x);  \label{10}
\end{equation}

\qquad

\begin{equation}
(\partial _{\mu }+n_{\mu }(n\partial ))A_{\mu }=-(n\partial )^{2}\Lambda
_{||},  \label{11}
\end{equation}%
\qquad

\ they are invariant relative coordinates inversion and changing sign at the
time reversing. As the commutator of $A_{\mu }^{(f)}$ obeys the Pauli-Jordan
function, additional fields form the nonlocal commutators:$\qquad $

\begin{equation}
\lbrack \Lambda _{0}(x),\Lambda _{0}(y)]=-[\Lambda _{||}(x),\Lambda
_{||}(y)]=iD_{N}(x-y),  \label{12}
\end{equation}%
\
\end{center}

Let's try to establish the reasons of appearance of this function.

The wave equation in vacuum is the operator record of dispersion identity $%
p^{2}\equiv \omega ^{2}-|k|^{2}=0$ and functions, that satisfy this
identity, are represented by the Fourier integral:$\qquad $

\begin{center}
\begin{equation}
G(x)=\int d^{4}p\ f(p)\ \delta (p^{2})e^{i(px)}  \label{13}
\end{equation}
\quad
\end{center}

with arbitrary, non-singular at $p^{2}=0$ function $f(p)$.

The partial solutions of corresponding non-uniform equation are represented
by the Fourier integral:\

\begin{center}
\begin{equation}
\overline{D}(x)=\frac{2}{(2\pi )^{4}}P\int d^{4}p\ \frac{1}{p^{2}}e^{i(px)}.
\label{14}
\end{equation}%
\quad
\end{center}

If to demand that (13) and (14) compose one analytical function that
corresponds to the Kramers-Kronig dispersion relations and to an opportunity
of energy transition from induced oscillations into free waves and back,
then $f(p)$ in (13) can consists of step operators $\theta (\pm \omega )$
only. If, however, such fields closely to the source are examining that does
not immediately generate far field waves, but can consist from the own or
confinement field of source only, this restriction on the form of $f(p)$ in
(13) loses its force (therefore the function $D_{N}$ had been appeared at
calculation of the electron self-field [9]). But it simultaneously means
that the acceleration of near field (e.g. at charge acceleration) should
lead to occurrence of free, far field waves, i.e.

\begin{center}
\begin{equation}
\partial _{t}^{2}D_{N}(x)\rightarrow D(x).  \label{15}
\end{equation}%
$\quad $
\end{center}

Hence the response functions of near field should be connected to the
corresponding Green functions (13) by the determination (possible constant
and linear on $t$ terms are omitted):\qquad

\begin{center}
\begin{equation}
D_{N}^{(.)}(x)=\frac{1}{2}\int\nolimits_{-\infty }^{\infty }d^{4}y\ n_{\mu
}|x_{\mu }-y_{\mu }|\ D^{(.)}(y),  \label{16}
\end{equation}%
\quad\
\end{center}

or, in the evident time-space form, as\
\begin{equation}
D_{N}^{(.)}(t,\mathbf{r})=\frac{1}{2}\int\nolimits_{-\infty }^{\infty }d\tau
\ |\tau -t|\ D^{(.)}(\tau ,\mathbf{r}),  \label{16'}
\end{equation}

that just corresponds the function, introduced by Schwinger.

The direct calculation, with the Green functions $D(t,r)$, $D^{(1)}(t,r)$
and $D^{(\pm )}(t,r)$ of wave equation in the right-hand side of (16'),
leads to such singular functions of near field:$\qquad $

\begin{center}
\begin{equation}
D_{N}(t,\mathbf{r})=\frac{1}{8\pi r}\{|t-r|\ -|t+r|\}\equiv \frac{1}{4\pi }%
\{sgn(t)\theta (t^{2}-r^{2})+\frac{t}{r}\ \theta (r^{2}-t^{2})\};  \label{17}
\end{equation}%

\begin{equation}
D_{N}^{(1)}(t,\mathbf{r})=\frac{1}{4\pi ^{2}r}\{(t+r)\ln (t+r)-(t-r)\ln
(t-r)\}\equiv \frac{1}{4\pi ^{2}}\{\ln (t^{2}-r^{2})-\frac{t}{r}\ln \frac{t\
-\ r}{t\ +\ r}\};  \label{18}
\end{equation}%

\begin{equation}
D_{N}^{(\pm )}(t,r)=\frac{1}{2}\{D_{N}(t,\ \mathbf{r})\mp iD_{N}^{(1)}(t,\
\mathbf{r})\},  \label{19}
\end{equation}
\end{center}

terms omitted in (16) can be restored for balancing dimensions of arguments
of logarithms in (18). Notice that $D_{N}(t,\ \mathbf{r})$ equals zero at $%
t=0$ just as $D(t,\ \mathbf{r})$.

It is remarkable that the temporal change of function (17) occurs completely
in the spatial direction:\

\begin{center}
\begin{equation}
\partial _{t}D_{N}(t,\ \mathbf{r})=\frac{1}{t}\ \mathbf{r\ \nabla }D_{N}(t,\
\mathbf{r})=-\frac{1}{4\pi r}\ \theta (r^{2}-t^{2}),  \label{20}
\end{equation}
\quad
\end{center}

which emphasizes the nonlocal character of near field in a concordance with
the tunnel character of (5').

Some features of these functions can be seen more obviously in the mixed $%
(\omega ,\mathbf{r})$-representation:\

\begin{center}
\begin{equation}
D_{N}(\omega ,\ \mathbf{r})=-\frac{1}{\omega ^{2}}D(\omega ,\ \mathbf{r})=-%
\frac{1}{2\pi i}\frac{\sin (\omega r)}{\omega ^{2}r};  \label{17'}
\end{equation}

\begin{equation}
D_{N}^{(1)}(\omega ,\mathbf{r})=\ sgn(\omega )D_{N}(\omega ,\ \mathbf{r});
\label{18'}
\end{equation}%

\begin{equation}
D_{N}^{(\pm )}(\omega ,\mathbf{r})=\theta (\pm \omega )D_{N}(\omega ,\
\mathbf{r}).  \label{19'}
\end{equation}%
\quad
\end{center}

So, in contrast to $D^{(.)}(\omega ,\mathbf{r})$, these functions are
singular at $\omega =0$.

Thus, the near zone of electromagnetic field, as can be asserting, is
nonlocal, in part at least, and is formed, in accordance with (12), by two
(additional) components of vector $A_{\mu }$ or by scalar fields
corresponding to their changes. (The opportunity of real replacement of
these field components by two scalar fields demands the in-depth analysis
and researches.)

But for all that the electric field $\mathbf{E}$, as must be emphasized,
remains local:\

\begin{center}
\begin{equation}
\left\langle T(E_{i}(x)E_{k}(y))\right\rangle =\partial _{x}\partial
_{y}\left\langle T(A_{i}(x)A_{k}(y))\right\rangle \rightarrow D(x-y).
\label{21}
\end{equation}%
\quad
\end{center}

A little differently the locality of $\mathbf{E}$ and $\mathbf{H}$ fields
can be shown by consideration of commutators:\

\begin{center}
\begin{equation}
\lbrack E_{i}(x),E_{j}(y)]=[H_{i}(x),H_{j}(y)]=\frac{1}{4\pi i}\{\partial
_{i}\partial _{j}-\delta _{ij}\partial _{t}^{2}\}D_{N}(x-y);
\end{equation}

\qquad

\begin{equation}
\lbrack E_{i}(x),H_{j}(y)]=\frac{1}{4\pi i}\partial _{t}\partial
_{j}D_{N}(x-y),  \label{22}
\end{equation}
\quad
\end{center}

double differentiation of the function $D_{N}(x)$ repays in $(\omega ,%
\mathbf{r})$ representation the factor $\omega ^{-2}$ responsible for
nonlocal effects in (6).

Thus, the nonlocality should be effective, in particular, in such $A_{\mu }$%
-depending phenomena as the Aharonov-Bohm effect, the Casimir effect, some
effects of near field optics, which become apparent at absence of electric
and magnetic fields and consequently without the Lorentz force.

\section{On quantum generalization of Lorentz condition}

The Lorentz-Fermi condition is written in QED as\

\begin{center}
\begin{equation}
\partial _{\mu }A_{\mu }^{(-)}\left\vert 0\right\rangle \equiv -iP_{\mu
}(0)\ A_{\mu }^{(-)}\left\vert 0\right\rangle =0,  \label{23}
\end{equation}
\quad
\end{center}

where $P_{\mu }(0)$ is the linear 4-momentum.

However the 4-moment dependence on a degree of inclusion of interaction (the
general adiabatic hypothesis) must be taken into account. In the covariant
Stueckelberg-Bogoliubov method [10] these features are described by the
interaction switching function (FIS) $g(x)\in \lbrack 0,1]$. So, in
particular, the 4-momentum depends on a degree of interaction switching as\

\begin{center}
\begin{equation}
P_{\mu }(g)=P_{\mu }(0)-\int d^{4}x\ H(x)\ \partial _{\mu }g(x),  \label{24}
\end{equation}%
\quad
\end{center}

$H(x)$ is the Hamiltonian of interaction.

Hence the condition (23) must be generalized as:\qquad

\begin{center}
\begin{equation}
-iP_{\mu }(g)A_{\mu }^{(-)}\left\vert g\right\rangle \equiv \partial _{\mu
}A_{\mu }^{(-)}\left\vert g\right\rangle -\int d^{4}y\ D^{(-)}(x-y)j_{\nu
}(y;g)\partial _{\nu }g(x)\left\vert g\right\rangle =0.  \label{25}
\end{equation}
\quad
\end{center}

The additional relation is executable at the Schwinger decomposition of
potential without a vacuum averaging . Therefore the performing of inner
part of (25) without averaging in near field can be \textit{assumed}:

\begin{center}
\begin{equation}
\partial _{\mu }A_{\mu }^{(-)}=\int d^{4}y\ D^{(-)}(x-y)\ j_{\nu }(y;g)\
\partial _{\nu }g(x).  \label{26}
\end{equation}%
\quad
\end{center}

In the particular system $n_{\mu }=\delta _{\mu 0}$ potentials of near field
are expressed by (11) as

\begin{center}
\begin{equation}
A_{0}^{(N)}(x)\equiv A_{0}-A_{0}^{(f)}=-\partial _{t}\Lambda _{0}(x);
\label{27}
\end{equation}%
\quad

\qquad

\begin{equation}
\mathbf{A}^{(N)}(x)\equiv \mathbf{A}-\mathbf{A}^{(f)}=-\mathbf{\nabla }%
\Lambda _{||}(x).  \label{28}
\end{equation}%
\quad
\end{center}

It shows that near field is non-vortex ($\mathbf{B}^{(N)}\equiv 0$), and its
electric component is pure longitudinal:\

\begin{center}
\begin{equation}
\mathbf{E}^{(N)}(x)\equiv -\partial _{t}\mathbf{A}^{(N)}-\mathbf{\nabla }%
A_{0}^{(N)}=\partial _{t}\mathbf{\nabla }(\Lambda _{0}+\Lambda _{||}).
\label{29}
\end{equation}
\quad
\end{center}

Scalar potential in the Coulomb gauge is equal zero, i.e. the field $\Lambda
_{0}(x)$ is stationary and does not take part in the definition (29).
Therefore in such gauge\qquad

\begin{center}
\begin{equation}
\mathbf{E}^{(N)}(x)\equiv -\partial _{t}\mathbf{\nabla }\Lambda _{||}=\int
d^{4}y[\partial _{t}\mathbf{\nabla }D_{N}^{(-)}(x-y)]\ j_{\mu }(y;g)\
\partial _{\mu }g(y)\equiv \mathbf{K}(x)\otimes _{x}J(x;g),  \label{30}
\end{equation}%
\quad
\end{center}

where $\mathbf{K}(x)=\partial _{t}\mathbf{\nabla }D_{N}^{(-)}(x)$ and $%
J(x;g)=j_{\mu }(y;g)\partial _{\mu }g(y)$ .

After inserting (20) the relation (30) represents the general expression of
near field strength and demonstrates the nonlocality of near field, nonzero
at any distance even at $t=0$.

As $D_{N}^{(-)}(x)=f(t,\ r)$, the formula (30) shows that near field is
determined by the radial component $E_{r}^{(N)}(x)$ only (the index $r$ is
below omitted). So in the $(\omega ,\mathbf{r})$-representation the kernel
of these forms is simplified:

\begin{center}
\begin{equation}
K(\omega ,\mathbf{r})=\frac{\theta (-\omega )}{2\pi i\omega }\partial _{r}%
\frac{\sin (\omega r)}{r},  \label{31}
\end{equation}%
\quad
\end{center}

and

\begin{center}
\begin{equation}
K(\omega ,\ \mathbf{k})=\frac{\theta (-\omega )}{(2\pi )^{3}i}\left\{ \frac{4%
}{(k\ -\ i0)^{2}\ -\ \omega ^{2}}-\frac{1}{\omega k}\ln \frac{k\ -\ \omega }{%
k\ +\ \omega }\right\} .  \label{32}
\end{equation}
\quad
\end{center}

Via the equation $\mathbf{\nabla E}^{(N)}(x)=4\pi \rho ^{(N)}(x)$, with
taking into account the wave equation, the expression (30) allows
determination of an effective space-time distribution of charges that forms
near field, i.e. allows the determination of dynamical form-factor of charge
system:\qquad

\begin{center}
\begin{equation}
\rho ^{(N)}(x)=\frac{1}{4\pi }\int d^{4}y\ j_{\mu }(y;g)[\partial _{\mu
}g(y)]\ \partial _{t}D^{(-)}(x-y)=\frac{1}{4\pi }\partial
_{t}D^{(-)}(x)\otimes _{x}J(x;g).  \label{33}
\end{equation}
\quad
\end{center}

Further analysis of (30) can be possible by substitution of the some
expressions of FIS' suggested in the Appendix.

\section{Near field of point-like charge}

Let's consider the fixed charge $Q$ in the origo of coordinates. Its
current density

\begin{center}
\begin{equation}
j_{\mu }(x)=Q\ \delta _{\mu 0}\delta (r)  \label{34}
\end{equation}
\quad
\end{center}

leads with taking into account the FIS (61) to the generalized current
function\

\begin{center}
\begin{equation}
J(x)=-\gamma Q\ \delta (r)sgn(t)e^{-\gamma |t|}.  \label{35}
\end{equation}
\quad
\end{center}

In the $(\omega ,\mathbf{r})$- representation

\begin{center}
\begin{equation}
J(\omega ,\mathbf{r})=\frac{i\gamma Q\omega }{\pi (\omega ^{2}\ +\ \gamma
^{2})}\ \delta (\mathbf{r})  \label{36}
\end{equation}
\quad
\end{center}

according to (30). It results in the following expressions for radial
components of electric field strength in $(\omega ,\mathbf{r})$- and $%
(\omega ,\mathbf{k})$-representations:\

\begin{center}
\begin{equation}
E^{(N)}(\omega ,\mathbf{r};\gamma )=\frac{\theta (-\omega )\gamma Q}{2\pi
^{2}(\omega ^{2}\ +\ \gamma ^{2})}\partial _{r}\frac{sin(\omega r)}{r},
\label{37}
\end{equation}
\quad

\

\begin{equation}
E^{(N)}(\omega ,\mathbf{k};\gamma )=\frac{\theta (-\omega )\gamma Q}{4\pi
^{3}(\omega ^{2}\ +\ \gamma ^{2})}\left\{ \frac{4\omega }{(k\ -\ i0)^{2}-\
\omega ^{2}}-\ln \left\vert \frac{k\ -\ \omega }{k\ +\ \omega }\right\vert
\right\} .  \label{37'}
\end{equation}
\quad
\end{center}

The inverse FT of (37) is expressed through the integral hyperbolic
functions (cf. [14], Exp. (108-9), $\beta =\gamma (t-r)$; $\overline{\beta }%
=\gamma (t+r)$):

\begin{center}
\qquad

\begin{equation}
E^{(N)}(t,\mathbf{r};\gamma )=\frac{2Q}{(2\pi )^{3}}\partial _{r}\left\{
\frac{1}{r}[i\pi e^{-|\beta |}+2\left\{ \text{chi}(|\beta |)\text{sh}(\beta
)-\text{shi}(\beta )\text{ch}(\beta )\right\} ]-\frac{1}{r}[\beta
\rightarrow \overline{\beta }]\right\} .  \label{38}
\end{equation}
\end{center}

Let's estimate the self energy of near field at the moment $t=0$:\

\begin{center}
\begin{equation}
W(0)=\frac{1}{8\pi }\int dr\ |E^{(N)}(t=0,\mathbf{r};\gamma )|^{2}.
\label{39}
\end{equation}%
\quad
\end{center}

Direct substitution of (38) into (39) leads to an excessively complicated
expression. Therefore we shall consider more rough estimations for FT of
(37) at $t=0$ for two frequencies regions taken separately:

\begin{center}
\begin{equation}
E_{1}^{(N)}(t=0,\mathbf{r}|\gamma >>\omega )\approx -\frac{Q}{\pi ^{2}\gamma
r^{3}};  \label{40}
\end{equation}%
\quad

\begin{equation}
E_{2}^{(N)}(t=0,\mathbf{r}|\gamma <<\omega )\approx -\frac{Q\gamma }{2\pi
^{2}r}.  \label{40'}
\end{equation}%
\ \quad
\end{center}

The expression (40) is related to low frequencies, i.e. to large distances,
and therefore (39) can be integrated in the limits $(1/\gamma ,\infty )$.
The expression (40') corresponds to high frequencies and after substitution
into (39) it can be integrated over $(0,1/\gamma )$. These integrations give
very close expressions and their sum leads to the compound estimation:

\begin{center}
\begin{equation}
W(0)\approx Q^{2}\gamma /6\pi ^{4}+Q^{2}\gamma /8\pi ^{4}\approx 3\cdot
10^{-3}Q^{2}\gamma  \label{41}
\end{equation}
\quad
\end{center}

or in the usual units and at $Q\rightarrow e$, $\alpha =e^{2}/\hbar c$, $%
[\gamma ]=\ $sec$^{-1}$\qquad
\begin{equation}
W(0)/mc^{2}\approx 3\cdot 10^{-3}\alpha \lambda _{C}\ \gamma /c=3\cdot
10^{-3}r_{0}\gamma /c.  \label{41'}
\end{equation}%
\quad

The estimation shows that on the distances of order of the Compton
wavelength, $\gamma \rightarrow c/\lambda _{C}$, the near field energy is of
order of some eV's that corresponds to potentials of ionization. At $%
c/\gamma $ of the order of Bohr radius, $r_{B}=\alpha ^{-1}\lambda _{C}$, it
gives \ decimal \ fractions of eV for near field energy\ in correspondence
with interatomic bonds.

The estimation of charge distribution of near field represents curious.
According to (231) and (37) in the $(\omega ,\mathbf{r})$-representation

\begin{center}
\begin{equation}
\rho ^{(N)}(\omega ,\mathbf{r})=-\frac{Q\omega ^{2}}{4\pi }g(\omega
)D^{(-)}(\omega ,\mathbf{r})=\theta (-\omega )\frac{iQ}{(2\pi )^{3}}\frac{%
\gamma \omega ^{2}}{\omega ^{2}\ +\ \gamma ^{2}}\frac{\sin \omega r}{r}.
\label{42}
\end{equation}
\end{center}

or%
\begin{equation}
\rho ^{(N)}(t,\ \mathbf{r})=\frac{1}{4\pi }\int d\tau \ g(\tau )\ \partial
_{t}^{2}D^{(-)}(\tau -t,r).  \label{42'}
\end{equation}
\quad

Its average value over $r$ or $t$ is equal, certainly, to zero, and the
maximum is achieved at $\omega r=0$. On high frequencies, when $\omega
>>\gamma $,

\begin{center}
\begin{equation}
\rho ^{(N)}(\omega ,\mathbf{r})\simeq Q\gamma D^{(-)}(\omega ,\mathbf{r}),
\label{43}
\end{equation}%
\quad
\end{center}

i.e. properties of near field come closer to far field features. In the low
frequencies field, i.e. at $\omega <<\gamma $, an effective charge density
in the near field $\rho ^{(N)}(\omega ,r)\rightarrow 0$.

\qquad

\section{Frustrated total internal reflection (FTIR)}

Let's consider, via the expression (30), the phenomenon of FTIR of light
wave $\mathbf{E}=\mathbf{E}_{0}e^{i\omega t}$ under an angle $\varphi $ ($%
\varphi >\varphi _{crit}$) from smooth dielectric surface ($z=0$) of medium
with polarizability $\alpha .$

The index of refraction is formally expressed as

\begin{center}
\begin{equation}
n(z)=n\ \theta (-z)+1\cdot \theta (z)=\frac{1}{2}[(n+1)-(n-1)sgn(z)].
\label{44}
\end{equation}%
\
\end{center}

In reality this idealized plane must be substituted by an effective layer of
minimal depth, without an instantaneous jump from parameters of one medium
to another ones, i.e. by a sufficiently smooth transitive layer between both
media. With this aim it is necessary to choose FIS, smooth together with the
first derivative, e.g. sgn$(z)\rightarrow g(z|\varphi )=\exp (-z^{2}/\Delta
z^{2})$ with $\Delta z^{-1}=4n(z)\omega \cos \varphi $.

The initial wave induces (or orients) dipoles $\mathbf{p}(t)=\alpha \mathbf{E%
}(t)$ on the surface \textquotedblleft plane\textquotedblright , i.e.
induces the "current" $j_{\mu }(y;g)\rightarrow (c/n)\delta _{\mu z}\ \rho
(t,\ \mathbf{r})$.

If to accept that a plane or layer interface of medium is strictly flat, the
density of surface charges would be described as

\begin{center}
\begin{equation}
\rho =\left\langle \mathbf{p}(t)\right\rangle \cdot \delta ^{\prime
}(z)\delta (x)\delta (y)=-\left\langle \mathbf{p}(t)\right\rangle \
z^{-1}\delta (z)\delta (x)\delta (y),  \label{45}
\end{equation}%
\quad\
\end{center}

i.e. as a double electric layer oscillating with a frequency of falling wave.

As this layer must be taken into account at estimations of other parameters
of medium also, it is necessary to replace $\delta (z)$ in the last
expression (45) on a $\delta $-like function in an agreement with the choice
of FIS: e.g. $\delta (z)\rightarrow $ $\delta (z,\xi )=(\xi \surd \pi
)^{-1}\exp (-z^{2}/\xi ^{2})$.

Thus the expression (30) of near field strength at inter-surface layer will
be of the order\

\begin{center}
\begin{equation}
\mathbf{E}^{(N)}(t,\mathbf{r}|\omega )=\alpha |\mathbf{E}_{0}|\int%
\nolimits_{-\infty }^{\infty }d\tau d\varsigma \ n^{-1}e^{-i\tau \omega
}\delta ^{\prime }(\varsigma ,\Delta z)[\partial _{\varsigma }g(\varsigma
|\Delta z)]\ \mathbf{\nabla }\partial _{\tau }D_{N}^{(-)}(\tau
-t;x,y,\varsigma -z),  \label{46}
\end{equation}
\quad
\end{center}

where the formal integration over time can be executed. It shows, that
frequencies of near field "photons", the evanescent "particles", will
coincide with frequencies $\omega $ of initial field and all their
differences should be manifesting only in momenta.

As $\partial _{z}g(z|\varphi )=-2z\Delta z^{-2}g(z|\varphi )$ and $\Delta
z(\varsigma )$ has different values for $\pm $\ arguments, the relation (46)
can be rewritten as

\begin{center}
\begin{equation}
\mathbf{E}^{(N)}(t,\mathbf{r}|\omega )=\frac{2i}{\sqrt{\pi }}\ \omega
e^{-i\omega t}\alpha |\mathbf{E}_{0}|\left\{ \left[ \frac{1}{\Delta z_{1}^{3}%
}\int\nolimits_{0}^{\infty }d\varsigma e^{(-2\varsigma ^{2}/\Delta
z_{1}^{2})}\mathbf{\nabla }D_{N}^{(-)}(\omega ;x,y,\varsigma -z)\right] +%
\frac{1}{n}\left[ \Delta z_{1}\rightarrow \Delta z_{2},\ z\rightarrow -z%
\right] \right\} ,  \label{47}
\end{equation}
\quad
\end{center}

where it is taken into account that $\Delta z\prime =n\Delta z$ does not
depend on $z$.

The distribution of near field, as shows (47), does not depend on depth of
its tunneling relatively FTIR "plane", in z-direction, and occurs
instantaneously. For this reason the possible transformation of evanescent
waves into extending waves occurs \textit{simultaneously} in all forbidden
depth (cf. [15]).

The expression (47) can be considered as the Fourier convolution over
variable $z$ in the first term and over $(-z)$ in the second one. By
separation of $z$-component of wave vector, $k=\{k_{\bot },q\}$ and with
taking into account the Fourier transformation (FT) of "current" factor:\

\begin{center}
\begin{equation}
I(q,\Delta z)=\int\nolimits_{0}^{\infty }d\varsigma e^{(iq\varsigma
-2\varsigma ^{2}/\Delta z^{2})}=\Delta z\sqrt{\frac{\pi }{8}}e^{-q^{2}\Delta
z^{2}/8}[1-\ erf(-iq\Delta z/\surd 8)].  \label{48}
\end{equation}%
\quad
\end{center}

So we receive the final expression for FT of near field intensity as\

\begin{center}
\begin{equation}
E^{(N)}(t,\mathbf{k}|\omega )=2\pi ^{-1/2}\omega e^{-i\omega t}\alpha
|E_{0}|\Delta z^{-3}k\{I(-q,\Delta z)+n^{-4}I(q,\Delta
z)\}D_{N}{}^{(-)}(\omega ;k).  \label{49}
\end{equation}%
\quad
\end{center}

At small values of parameter $q\Delta z$ the expression in braces in (49)
becomes

\begin{equation}
\{\ldots \}\rightarrow \sqrt{\frac{\pi }{8}}\frac{1}{\Delta z^{2}}[\exp
(-q^{2}\Delta z^{2}/8)+\frac{1}{n^{3}}\exp (-n^{2}q^{2}\Delta z^{2}/8)],
\label{50}
\end{equation}

i.e. it contains only $q$-components of wave vector of any magnitude.
Therefore "evanescent photons" can possess, at the same frequency, a bigger
or smaller momenta (compare [16]).

But for all that, due to the identity of frequencies, they can interfere
with photons of inlet radiation. It means that such interference can locate
objects, at supervision of interference picture closely enough to the
refraction surface, with sizes about $|k_{z}-q|^{-1}$, i.e. smaller light
wavelengths in vacuum. Just this effect is the physical basis of so-called
near field optics [4] (we do not examine the evident complications
associated with light polarization).

Let's note that the use of another FIS, e.g. (67) instead of (65), leads to
similar results with replacement $q^{2}\Delta z^{2}$ on $|q\Delta z|$ and
insignificant change of numerical factors. The choice between different FIS'
can be made, at the given stage, by a comparison with experiments only.

\qquad

\section{Interaction of atoms in near field}

\textbf{\ }

Energy of non-resonant interaction of two neutral atoms on distances smaller
wavelength, but bigger their own sizes [11], is determined by the two-photon
exchange (the fourth order of $S$-matrix) as\

\begin{center}
\begin{equation}
U(\mathbf{r})=\frac{i}{4\pi }\int\nolimits_{-\infty }^{\infty }d\omega \
\omega ^{4}\alpha _{1}(\omega )\alpha _{2}(\omega )[D_{ik}(\omega ,\mathbf{r}%
)]^{2},  \label{51}
\end{equation}
\quad
\end{center}

where $\alpha _{i}(\omega )$ is the polarizability of cooperating atoms,
scalar at the $S$-states.

The affinity of $D_{il}(\omega ,r)|_{NF}$ to the matrix element of
dipole-dipole interaction is evident. Therefore for such distances and
atomic frequencies that $\omega r<<1$, the calculations with inserting the
decomposition (3) into (51) are precisely identical to the procedure [11]
and lead to the Van-der-Waals energy of interaction proportional $R^{-6}$
and to the Casimir energy of interaction of atoms proportional to $R^{-7}$.

Thus, these interactions are describable by the propagator (6), i.e. they
occur in the near field and, at least, are in part transferred
superluminally.

However for resonant interaction between identical (motionless) atoms,

\begin{center}
\begin{equation}
A_{1}^{\ast }+A_{2}\longleftrightarrow \ A_{1}+A_{2}^{\ast },  \label{52}
\end{equation}
\quad
\end{center}

matrix element is nonzero still in the second order:

\begin{center}
\begin{equation}
S^{(2)}=-\frac{1}{2}\int dt_{1}dt_{2}\ T\{V(t_{1})V(t_{2})\},  \label{53}
\end{equation}%
\
\end{center}

where $V=-\mathbf{E}(\mathbf{r}_{1})\mathbf{d}_{1}-\mathbf{E}(\mathbf{r}_{2})%
\mathbf{d}_{2}$. Therefore instead of (51) we have

\begin{center}
\begin{equation}
U(\mathbf{r})=(i/2\pi )\int\nolimits_{-\infty }^{\infty }d\omega \ \omega
^{2}D_{ik}(\omega ,\mathbf{r})\ Re[\alpha _{ik}(\omega )],  \label{54}
\end{equation}%
\
\end{center}

with the tensor of scattering of two-level, for simplicity, systems
expressed through matrix elements of dipole momenta:

\begin{center}
\begin{equation}
\alpha _{ik}(\omega )=\frac{(d_{i})_{01}(d_{k})_{10}}{\omega _{0}\ -\ \omega
\ -\ i\Gamma }+\frac{(d_{k})_{01}(d_{i})_{10}}{\omega _{0}\ +\ \omega \ -\
i\Gamma }.  \label{55}
\end{equation}
\end{center}

By the substitution of $D_{N}$ function into (54) it can be shown that the
interaction (52) decreases as $R^{-3}$ (cf. [17]).

The full probability of process (52) in the near field is determined as\

\begin{center}
\begin{equation}
W\ \propto \int\nolimits_{-\infty }^{\infty }d\omega |D_{ik}(\omega ,\mathbf{%
r})\alpha _{ik}(\omega )|^{2}\rightarrow \int\nolimits_{-\infty }^{\infty
}d\omega |d_{1}|^{2}|d_{2}|^{2}|D_{ik}(\omega ,\mathbf{r})|_{NF}|^{2}\tau
(\omega )/\Gamma ,  \label{56}
\end{equation}%
\quad
\end{center}

where the expression of duration of scattering process is separating out:

\begin{center}
\begin{equation}
\tau (\omega )=\Gamma /2[(\omega _{0}-\omega )^{2}+\Gamma ^{2}/4].
\label{57}
\end{equation}%
\qquad
\end{center}

By carrying out integration in view of $\delta $-character of (57), using
matrix elements of dipole operators $|d|^{2}=\hbar e^{2}f/2m\omega $, $f$ is
the oscillator force, and substituting the expressions of singular functions
(6-7), we receive, that the probability of process depends on distance
between cooperating atoms as $R^{-6}$, i.e. it takes the form of the
well-known half-empirical F\"{o}rster law [18] (see, e.g., [19]):

\begin{center}
\begin{equation}
W=\Gamma ^{-1}(R_{0}/R)^{6},  \label{58}
\end{equation}%
\
\end{center}

where $R_{0}$ is the so-called F\"{o}rster radius.

With (6) it follows that the rate of process (52) in the time representation
is represented by the square of near field singular function (6):

\begin{center}
\begin{equation}
|D_{il}(t,\ \mathbf{r})|_{NF}|^{2}=\frac{1}{(4\pi r^{2})^{2}}\left\{ \theta
(t^{2}-r^{2})+(\frac{t}{r})^{2}\theta (r^{2}-t^{2})\right\} ,  \label{59}
\end{equation}%
\
\end{center}

that determines relative probabilities of excitation transfer with
subluminal and superluminal speeds.

But the superluminal (instantaneous) interaction is possible, as was
established in [7], at the tunneling only. Are there certain processes and
certain frequencies ranges in condensed media similar to anomalous
dispersion in optics?

In the articles [20] we had shown that at phase transitions of the first
kind the liberated latent energy must be converted, at least partially, into
the characteristic radiation with frequencies determined by released energy
at establishing definite bonds. (This phenomenon is hard for observations,
because surrounded substances quickly thermalize the emitted radiation, but
its existence is confirmed by some experiments, e.g. [21] and references
therein.)

This phenomenon can take place at the sight of discrepancy between energy of
relating corresponding bonds and momenta of emitted virtual "photons".
Therefore their interaction with neighbors can have some similarity with
anomalous dispersion regime, in particular with its instantaneous
peculiarity. It allows to \textit{assume} that the permanent interatomic
bonds in condensed state, partially or completely, are instantaneous ones.

\section{Conclusions}

Let us enumerate the results.

1. The decomposition of canonical Green functions of QED leads to the
appearance of propagators of far and near fields and an intermediate one
describing transitions between them.

2. Green function of near field corresponds to the Schwinger function,
initially introduced for investigation of an electron self-field. The rule
of transition from far field functions into near field ones and vice versa
is considered.

3. The Schwinger scheme of $A_{\mu }$ decomposition allows to discard at
examination of near field phenomena a formal vacuum averaging of the
classical Lorentz condition. At the same time it shows that the near field
function is formed by longitudinal and scalar (temporal) components of the
4-vector $A_{\mu }$ or by two additional scalar fields, derivatives of which
represent these components. Thus this scheme demonstrates the physical sense
of "surplus" components and shows the inconsistency of their complete formal
elimination by introduction of indefinite metrics.

4. The performed analysis has shown, within the frame of QED, that the near
zone of $A_{\mu }$ represents the nonlocal, but quickly decreasing field, so
that the $\mathbf{E}$ and $\mathbf{B}$ fields remain local. Hence this
analysis requires the introduction of the \textquotedblright nonlocality in
the small\textquotedblright\ only.

It once more underlined the necessity and significance of notions of
adiabatical switching on and off interaction for understanding details of
QED interactions (cf., e.g. [22]).

5. The general approach to FTIR phenomena is elaborated via \textit{%
introduction}, along an analogue with the field theory, the function of
interaction switching at wave transition into another medium. Such method
can be applied to other phenomena of substance parameters changing. The used
approach shows the superluminal features of FTIR.

6. Instantaneous or, more widely, superluminal transfer of excitations can
be or even must be detectable in other phenomena. In this connection it is
shown that the common expressions for the Van-der-Waals potential and the
Casimir forces can be naturally expressed through the Schwinger function. It
means that these interactions can be, partly at least, instantaneous. The
expression for excitation transfer on small distances (the F\"{o}rster law)
also has the same form.

Let's recall, in this connection, the continued discussions of the temporal
features of tunneling [23] that had induced a number of paradoxes (e.g.,
[24, 25]). Our consideration shows within the framework of QED, at least,
that the tunnel transition must be executable within the scope of near field
and, under some conditions, can be instantaneous [7].

All our consideration shows that the instantaneous transferring is not of
very exotic, extraordinarily nature; its manifestations can occur in some
phenomena, which may be considered as the "nonlocal in the small", and
therefore their temporal features should be investigated more carefully and
widely. (It seems, for example, that such phenomena as the energy-time
entanglements can be also connected with such nonlocality, e.g. [27].)

7. The described features of near zone of point-like charge and, on the
other hand, of optical transitive zones as the fields of longitudinal and
scalar evanescent "photons" with possibilities of observability of their
"superluminal" features, deprives the formal schemes of elimination of
"superfluous" fields components, such as introduction the indefinite
metrics, their general significance. All this demands anew returning to the
principal problems of the QED gauges, opportunities of their transformations
and their peculiarities.

8. The revealed "nonlocality in the small" in the context of covariant field
theory requires not only more scrupulous further research of its properties,
but also the search of similar phenomena as in the QED, so, probably, in the
theories of other fields.

\section{\textbf{A}cknowledgments}

The author is grateful to Professors F. W. Hehl, G. Nimtz, I. I. Royzen and
Dr. G. M. Rubistein for valuable comments and support.

\section{Appendix: Functions of interaction switching}

\textbf{\ }

In the Bogoliubov theory the temporal FIS' $g(t,\ \mathbf{r})$ are not
concretized, they must only satisfy the conditions:

a.\qquad $g(t,\mathbf{\ r})\in \lbrack 0,1]$, which can be slightly
generalized as $|$$g(t,\ r)|\ \in \lbrack 0,1]$;

b.\qquad General covariance;

c.\qquad Limiting conditions: $g(x)\rightarrow 0$ at $x\rightarrow \pm
\infty $;

d.\qquad $g(-x_{\mu })=g(x_{\mu })$, that follows CPT invariance of $A_{\mu
} $.

The simplest FIS' satisfying these conditions:

\begin{center}
\begin{equation}
g_{1}(x)=e^{-\gamma |n_{\mu }x_{\mu }|};\quad g_{2}(x)=e^{-\gamma
^{2}(n_{\mu }x_{\mu })^{2}};\quad \ g_{3}(x)=e^{-\gamma ^{2}|x_{\mu }|^{2}}.
\label{60}
\end{equation}%
\
\end{center}

\qquad The existence of near field in classical theories allows its
independence from $\hbar $. Therefore for classical problems it seems
natural to express this parameter by the Thompson radius of electron: $%
\gamma =c/r_{0}\equiv mc^{3}/e^{2}$, this form leads to disappearance of
interaction and near field at $e\rightarrow 0$. For bound electron its
expression via the Compton wavelength of electron seems natural: $\gamma
^{\prime }=c/\lambda _{C}=mc^{2}/\hbar $.

For the part of our consideration the choice $n_{\mu }=\delta _{\mu 0}$ is
sufficient:

\begin{center}
\begin{equation}
g_{1}(x)=e^{-\gamma |t|};\quad g_{2}(x)=e^{-\gamma ^{2}t^{2}}  \label{61}
\end{equation}%
\qquad\
\end{center}

and $g_{2}(x)=g_{3}(x)$ at $r=0$. Under the FT these FIS' take the forms:

\begin{center}
\begin{equation}
g_{1}(\omega )=\gamma /\pi (\omega ^{2}+\gamma ^{2});\quad g_{2}(\omega
)=(2\gamma \surd \pi )^{-1}\exp (-\omega ^{2}/4\gamma ^{2}),  \label{62}
\end{equation}%
\qquad\ \
\end{center}

at $\gamma \rightarrow 0$ they aspire to $\delta (\omega )$.

To functions (62) different physical interpretations can be given. So, $%
g_{2}(\omega )$ corresponds to the normal law of probability with similar
process of interaction switching, etc. The function $g_{1}(\omega )$
corresponds, excluding factor $\pi $, to the duration of elastic photon
scattering on free electron. Such interpretation allows generalization of $%
g_{1}(\omega )$ for interaction with bound electron into two level systems:

\begin{center}
\begin{equation}
g_{1}(\omega |\omega _{0})=\gamma /2\pi \{[(\omega -\omega _{0})^{2}+\gamma
^{2})]^{-1}+[(\omega +\omega _{0})^{2}+\gamma ^{2})]^{-1}\}  \label{63}
\end{equation}%
\
\end{center}

or, in the $t$-representation,

\begin{center}
\begin{equation}
g_{1}(t|\omega _{0})=e^{-\gamma |t|}\cos (\omega _{0}t).  \label{64}
\end{equation}%
\
\end{center}

Let us introduce on the similar basis the space FIS' determining the
switching or alteration of interaction during particle (wave) transitions
across determined (flat) space borders. So, at approach of light wave to
refracting surface $z=0$ it can be \textit{suggested} the FIS:

\begin{center}
\begin{equation}
g_{1}(x)\rightarrow g_{1}(z)=e^{-\kappa |z|};\quad g_{2}(x)\rightarrow
g_{2}(z)=e^{-\kappa ^{2}z^{2}},  \label{65}
\end{equation}%
\
\end{center}

where $\kappa ^{-1}$ is the effective width of an intermediate surface layer
depending on parameters of both substances and light flux.

At the total internal (or external) reflection of light of frequency $\omega
$ on intersection of media with indices of refraction $n_{1}$ and $n_{2}$
under angle $\varphi >\varphi _{crit}$, $z$-component of photon momentum $%
k_{z}=kn_{1}\cos \varphi $ changes sign and the alteration of momentum is
equal

\begin{center}
\begin{equation}
|\Delta k_{z}|\ =2kn_{1}\cos \varphi .  \label{66}
\end{equation}%
\
\end{center}

In accordance with the uncertainty principle this alteration is executed in
the layer $|\Delta z|\ =1/2|\Delta k_{z}|$, therefore the value $\kappa
=1/|\Delta z|$ and at $n=n_{1}/n_{2}$

\begin{center}
\begin{equation}
g_{1}(t,r|\varphi )=e^{-4n|\omega ||z|\cos \varphi },\qquad \quad
g_{2}(t,r|\varphi )=e^{-(4n\omega z\cos \varphi )^{2}}.  \label{67}
\end{equation}%
\
\end{center}

Such transient optical layer should exist at all processes of reflection and
refraction, since at $\varphi <\varphi _{crit}$ and with refraction angle $%
\varphi ^{\prime }$ the change of photon momentum is

\begin{center}
\begin{equation}
|\Delta k_{z}|=k|n_{1}\cos \varphi -n_{2}\cos \varphi \prime |,\   \label{68}
\end{equation}
\end{center}

which again leads to (67).\

These processes, as it is known, can be of nonlocal character, described
usually via surface polaritons (e.g., [28]).

\subsection{References}

*). E-mail: mark\_perelman@mail.ru\qquad

1. E. Fermi. Rev.Mod.Phys., \textbf{4}, 87 (1932).

2. G. K\"{a}llen. \textit{Quantum Electrodynamics}. London: 1972.

3. Y. Aharonov, D. Bohm. Phys.Rev., \textbf{115}, 485 (1959).

4. C. Girard and A. Dereux. Rep. Prog. Phys. \textbf{59}, 657 (1996); C.
Girard, C. Joachim and S. Gauthier. Rep. Prog. Phys. \textbf{63}, 893 (2000).

5. O. Keller. J.Opt.Soc.Am.B, \textbf{16}, 835 (1999); Phys.Rev.A, \textbf{62%
}, 022111 (2000).

6. G. Nimtz and W. Heitman. Progr.Quant.Electr., \textbf{21}, 81 (1997); R.
Y. Chiao and A. M. Steinberg. Phys.Scr.T, \textbf{76}, 61 (1998); E. Recami.
Found.Phys., \textbf{31}, 1119 (2001); P. W. Milonni. J.Phys.B., \textbf{35}%
, R31 (2002); G. Nimtz. Progr.Quant.Electr., \textbf{27}, 417 (2003).

7. M. E. Perel'man. Ann. Phys. (Leipzig), \textbf{14}, 733 (2005).

8. M. E. Perel'man. Sov. Phys. Doklady, \textbf{19}, 26 (1974); Int. J.
Theor.Phys. - in press; quant-ph/0510123.

9. J. Schwinger. Phys.Rev., \textbf{75}, 651 (1949), \textbf{74}, 1439
(1948).

10. N. N. Bogoliubov and D. V. Shirkov, \textit{Introduction to the Theory
of Quantized Fields}, 3rd edition (Wiley, New York, 1980).

11. V. Berestetski, E. Lifshitz and L. Pitayevski, \textit{Relativistic
Quantum Theory} (Pergamon, Oxford, 1971).

12. L. D. Landau and E. M. Lifshitz, \textit{The Classical Field Theory}
(Pergamon, Oxford, 1975).

13. E. Wolf and H. T. Foley. Opt.Lett., \textbf{23}, 16 (1998).

14. Ya. Brychkov and A. Prudnikov, \textit{Integral Transformations of
Generalized Functions} (Moscow: Nauka, 1977).

15. K. Wynne and D. A. Jaroszynski. Opt.Lett., \textbf{24}, 25 (1999);
D.R.Solli e.a. Phys.Rev.Lett, \textbf{91}, 143906 (2003).

16. J. M. Vigoureux, L. D'Hooge and D. Van Labeke. Phys.Rev.A, \textbf{21},
347 (1980).

17. \textit{Long Range Forces: Theory and Recent Experiments in Atomic
Systems }(F. S. Levin, D. Misha, Ed's). NY, Plenum, 1992.

18. Th. F\"{o}rster. Ann. Phys. (Leipzig), \textbf{2}, 55 (1948).

19. S. Y. Buhmann e.a.: quant-ph/0403128.

20. M. E. Perel'man. Phys.Lett. A, \textbf{32}, 411 (1971);
Sov.Phys.-Doklady, \textbf{19, }26 (1974); Astrophysics, \textbf{17}, 383
(1981); arXiv: cond-mat/0401544, Phil. Mag. (2006) - in press.

21. L. M. Umarov and V. A. Tatarchenko. Crystallography, \textbf{29}, 1146
(1984).

22. P. W. Milonni, R. J. Cook and J. R. Ackarhalt. Phys.Rev.A, \textbf{40},
3764 (1989).

23. L. A. MacColl. Phys.Rev., \textbf{40}, 621 (1932).

24. E. O. Kane. In: \textit{Tunneling Phenomena in Solids}. (E.Burstein,
S.Lundqvist, Ed's). NY: Plenum, 1969.

25. F. E. Low and P.F. Mende, Ann. Phys. (NY) \textbf{210}, 380 (1991); M.
Morgenstern et al. Phys.Rev.B, \textbf{63}, 201301 (2001).

26. J. Broe and O. Keller. Opt.Comm. \textbf{194}, 83 (2001) and references
therein.

27. S. Faisel et al. Phys.Rev.Lett., \textbf{94}, 110501 (2005).

28. A. Zangwill, \textit{Physics at Surfaces} (Cambridge Univ. Press,
London, 1988).

\end{document}